\documentclass[aps,prl,onecolumn,showpacs,preprintnumbers]{revtex4}
\usepackage{amsmath}
\usepackage{dcolumn}
\usepackage{bm}
\usepackage{subfigure}
\usepackage{amsfonts}
\usepackage{amssymb}
\usepackage{makeidx}
\usepackage{epsfig}
\usepackage{graphicx}

\begin{document}

\title{\textbf{Carter-constant induced mechanism for generation of
anisotropic kinetic equilibria in collisionless }$N-$\textbf{body systems}}
\author{Claudio Cremaschini\thanks{%
Electronic-mail: claudiocremaschini@gmail.com}$^{a}$ and Zden\v{e}k Stuchl%
\'{\i}k$^{a}$}
\affiliation{$^{a}$Institute of Physics and Research Center for Theoretical Physics and
Astrophysics, Faculty of Philosophy and Science, Silesian University in
Opava, Bezru\v{c}ovo n\'{a}m.13, CZ-74601 Opava, Czech Republic}
\date{\today }

\begin{abstract}
A new intrinsically-relativistic kinetic mechanism for generation of
non-isotropic relativistic kinetic equilibria in collisionless $N-$body
systems is pointed out. The theory is developed in the framework of the
covariant Vlasov statistical description. The new effect is based on the
constraints placed by the conservation laws of neutral single-particle
dynamics in prescribed background curved-spacetimes demonstrating existence
of Killing tensors. As an illustration, the particular case of the Kerr
space-time admitting the so-called Carter constant for the particle geodesic
motion is considered. The general functional form of the equilibrium kinetic
distribution function (KDF) is determined and an explicit realization in
terms of Gaussian-like distributions is provided. It is shown that, due to
the Carter constant, these equilibrium KDFs exhibit an anisotropic
phase-space functional dependence in terms of the single-particle 4-velocity
components, giving rise to corresponding non-isotropic continuum fluid
fields. The qualitative properties of the equilibrium stress-energy tensor
associated with these systems are discussed, with a particular emphasis on
the related occurrence of temperature anisotropy effects. The theory is
susceptible of astrophysical applications, including in particular the
statistical properties of dark matter halos around stellar-mass or
galactic-center black holes.
\end{abstract}

\maketitle

\section{1 - Introduction}

The subject of the present research belongs to statistical mechanics
developed in the framework of General Relativity. More precisely, the paper
deals with the search of possible intrinsically general-relativistic effects
that can be responsible for the occurrence of non-isotropic kinetic
distributions in collisionless $N-$body systems, namely characterized by
generally species-dependent kinetic distribution functions (KDFs) $f_{%
\mathrm{j}}$ $\equiv f_{\mathrm{j}}(\mathbf{x},s)$ which are non-isotropic
with respect to the single-particle $4-$velocity components $u^{\mu }$. Here
$s$ and $\mathbf{x}\equiv \left( r^{\mu },u^{\mu }\equiv dr^{\mu }/ds\right)
$ denote respectively the single-particle proper time and state, with $%
\mathrm{j}$ being the species index.\textbf{\ }In statistical physics the
KDF is defined in the single-particle phase-space $x$, namely velocity and
configuration spaces, and has a statistical meaning, representing the
fundamental quantity in terms of which the physical observable properties of
a system (i.e., the continuum fluid fields, like density and temperature)
can be determined. The KDFs may correspond, in particular, to the so-called
kinetic equilibria \cite{PoP2014-1,PoP2014-2}.\textbf{\ }This means that,
when referred to a suitable general-relativistic reference frame, all the
species KDFs $f_{\mathrm{j}}$ are required to be smooth, strictly-positive
ordinary functions which depend only on a suitable set of invariants $I_{l}(%
\mathbf{x})$, namely such that $\frac{d}{ds}I_{l}(\mathbf{x}(s))=0$
identically,\ where $\mathbf{x}\left( s\right) $\ is the single-particle
phase-space trajectory parametrized in terms of the particle proper-time $s$%
. More precisely, here the real scalar functions $I_{l}(\mathbf{x})$\
identify the so-called integrals of motion, i.e., invariant phase functions
which do not explicitly depend on $s$ \cite{wald} (see also definition
below).

The importance of the subject lays both in the actual realization of a
possible new type of non-isotropic relativistic kinetic equilibria and the
identification of the physical effects of anisotropy emerging at the level
of a fluid statistical description, i.e., achieved via suitable fluid tensor
fields - in particular the stress-energy tensor - which are prescribed by
means of appropriate $4-$velocity moments of the equilibrium KDF \cite%
{degroot}. In this context, the anisotropy means that the continuum fluid
described by the statistical kinetic theory can be characterized by
macroscopic observable properties exhibiting different profiles depending on
the spatial direction along which these are evaluated. The most
representative example of such a behavior is given by the occurrence of
temperature and/or pressure anisotropies. On general grounds one can say
that phase-space anisotropies leading to such effects might arise whenever
the underlying KDF of the system deviates from an isotropic Maxwellian\
distribution (see detailed discussion reported below). The focus of the
investigation concerns the treatment of collisionless $N-$body systems
composed by neutral matter arising in the curved space-time surrounding
astrophysical compact objects and the search of equilibrium KDFs which are
associated, in such environments, with the conservation laws of
single-particle dynamics. Because of the nature of the new mechanism
proposed here, the present theory is susceptible of astrophysical
applications in disparate contexts, i.e.,\ characterized by a wide range of
the relevant physical parameters. Among the most relevant ones, we mention
the case of collisionless neutral gas clouds \cite%
{gas0,gas00,gas1,gas2,gas3,gas4} and dark matter (DM)\ halos \cite%
{DM1,DM3,DM4,DM5,DM6,DM7,DM8} around stellar-mass or galactic-center black
holes, as well as possibly the case of collisionless stellar systems
belonging to globular clusters or orbiting central galactic regions \cite%
{gravi,r85,star0,star1}.

Different statistical descriptions can, in principle, be implemented for
large $N-$body systems, namely which are formed by a number $N$ of
constituent particles which is considered $\gg 1$\ but still finite. These
include essentially two types of approaches: A) relativistic microscopic and
kinetic theories \cite{degroot,gravi,PoP2014-2,EPJ1,EPJ2,EPJ5}, which are
built on particle dynamics and conservation laws and retain information
about the statistical distribution of matter in particle configuration and
velocity spaces through the KDF; B) fluid theories, i.e., relativistic
hydrodynamic (HD) \cite{ros,mano2,rezhd2} and/or magnetohydrodynamic (MHD)
\cite{rezmhd,ma-2,ma-3,ma-4,ma-5,ma-6} treatments, according to which the $N-
$body system is treated as a continuum in configuration space and is
characterized by suitable observable fluid fields. The choice of the
treatment depends of course on the type of relevant physical phenomena to be
investigated, i.e., whether related to the multi- or single-particle
dynamics (case A) or depending on their \textquotedblleft
macroscopic\textquotedblright\ dynamics described by means of a complete set
of fluid fields (case B), for neutral or electrically-charged matter. It
must be stressed, however, that the completeness feature mentioned here is a
mandatory prerequisite for the validity of all fluid theories. This requires
the precise knowledge of the fluid equations evolving the fluid fields of
the system (e.g., mass and energy density, temperature and velocity fields)
to be completed by the prescription of the consistent fluid closure
conditions (e.g., stress-energy or pressure tensors) \cite%
{coll-2011,coll-2014}.

Nevertheless, in the relativistic and astrophysical context which is
considered here both single-particle and macroscopic fluid $4-$velocities
are by assumption intrinsically relativistic in character, mainly because of
space-time curvature effects associated with strong gravitational fields.
Under these circumstances,\ either microscopic or kinetic relativistic
statistical approaches need to be adopted. In particular, in the case of
collisionless multi-species systems, the appropriate statistical framework
is provided by the covariant Vlasov kinetic theory, to be generally
supplemented by both Maxwell and Einstein equations. This type of kinetic
description in fact allows for both phase-space single-particle as well as
(electromagnetic and gravitational) collective system dynamics to be
consistently taken into account \cite{degroot}. Within such a description,
the fundamental quantity is represented by the species KDF $f_{\mathrm{j}}$.
The KDF is defined in the single-particle phase-space and its dynamical
evolution is determined by the Vlasov equation. In Lagrangian form, the
latter is written as%
\begin{equation}
\frac{d}{ds}f_{\mathrm{j}}\left( \mathbf{x}\left( s\right) ,s\right) =0,
\label{vlasov-1}
\end{equation}%
where\ in general the function $f_{\mathrm{j}}$\ can still depend explicitly
on $s$. The velocity-integrals of the KDF define appropriate fluid fields,
namely physical observables, while the velocity moments of the kinetic
equation (\ref{vlasov-1}) determine the corresponding set of continuum fluid
equations. In this sense the kinetic approach is more fundamental than the
fluid one. In addition, it must be stressed that, in the framework of the
kinetic approach, the prescription of the closure conditions for the fluid
equations becomes unique and self-consistent once a solution of the Vlasov
equation is obtained. The issue is particularly relevant in collisionless
systems where kinetic effects must be properly retained in the physical
description, implying the fluid closure conditions to become non-trivial
\cite{APJS}. This originates from the fact that collisionless $N-$body
systems can develop phase-space anisotropies, both in equilibrium as well as
in non-stationary configurations, which would be otherwise scarsely
understandable by a fluid treatment ignoring the underlying kinetic
properties. Effects of this type can arise either:\newline
A) Due to relativistic single-particle dynamics and related conservation
laws \cite{p-1,p-2,p-3,p-4,p-5}, including the occurrence of adiabatic
invariants \cite{p-0}. For charged particles in electromagnetic fields these
features are enriched by gyrokinetic dynamics in plasmas \cite%
{PoP2014-1,PoP2014-2,Bek1,Bek2} and plasma kinetic regimes \cite{Cr2012},
taking into account the constraints placed by non-uniform magnetic fields on
particle dynamics and confinement mechanisms.\newline
B) Due to the peculiar form of the KDF itself, which may depart from an
isotropic local Maxwellian function \cite{Cr2010,Cr2011} (see also Section 3
below), implying the occurrence of collective phenomena which are
distinctive of the statistical behavior of collisionless $N-$body systems.

The statistical treatment of these features is essential in order to
characterize the dynamical and thermodynamical properties of the systems of
interest. For example, phase-space anisotropies of the type mentioned above
can be responsible for the occurrence of collective drift-velocities, namely
velocity components of kinetic origin having peculiar spatial directions and
which affect the bulk motion of the continuum fluid introducing drift
corrections. In the case of plasmas different mechanisms have been
identified which are responsible for the appearance of such drifts. The most
relevant ones are:\ 1) those associated with conservation of particle
canonical momentum in axisymmetric systems in the presence of strong
magnetic fields, i.e. the so-called diamagnetic effects in
strongly-magnetized plasmas; 2)\ those associated with Larmor rotation of
charges and the related confinement mechanism for particle trajectories
approximating magnetic field lines in magnetized plasmas, usually denoted as
finite Larmor-radius (FLR) effects;\ 3)\ those associated with energy
conservation in stationary or slowly time-varying kinetic equilibria \cite%
{APJS,Cr2010,Cr2011}. In addition, phase-space anisotropies can generate
also shear-flow phenomena characterized by strong gradients of angular
velocity in axisymmetric systems \cite{Cr2013c} as well as magnetic field
generation (kinetic dynamo)\ by local current densities in plasmas \cite%
{coll-2011,Cr2011a}. Similarly, temperature and pressure anisotropies as
well as non-vanishing heat fluxes can arise on this basis.

For non-relativistic plasmas in astrophysical and laboratory contexts, these
issues have been investigated systematically in a series of recent works
(see for example Refs.\cite{Cr2010,Cr2011}), where kinetic equilibria were
constructed in the framework of the Vlasov-Maxwell description, together
with the establishment of absolute stability criteria in the presence of
axisymmetric electromagnetic perturbations \cite{PRL,PRE-new}. Remarkably,
the theory was proved to apply to axisymmetric configurations, spatially
non-symmetric kinetic equilibria in which energy is conserved \cite{Cr2013}
as well as energy-independent kinetic equilibria \cite{PRE-new} in which a
continuous spatial symmetry of some kind still survives but the KDF does not
carry a dependence on the particle energy. Different physical sources of
velocity-space anisotropies for the KDF have been pointed out in plasmas,
which determine correspondingly non-isotropic pressure tensors. The most
relevant one is enhanced by the adiabatic conservation of the guiding-center
particle magnetic moment as predicted by gyrokinetic theory, which is
associated with the Larmor rotation in non-uniform magnetic fields. This
anisotropy was shown to contribute also generating at equilibrium
non-vanishing species flow velocities and driving a kinetic dynamo mechanism
in combination with FLR-diamagnetic effects and non-uniform plasma fluid
fields \cite{coll-2011,Cr2011a}.

The same type of physical mechanism is expected to operate also for plasmas
in curved space-time, as suggested by the relativistic kinetic theory
established in recent contributions \cite{PoP2014-1,PoP2014-2}. In
particular, in Ref.\cite{PoP2014-1} kinetic equilibria of relativistic
collisionless plasmas in the presence of non-stationary electromagnetic
fields have been addressed, while Ref.\cite{PoP2014-2} dealt with the
covariant formulation of spatially non-symmetric kinetic equilibria in
magnetized plasmas and the determination of the physical mechanisms
responsible for the occurrence of a non-vanishing $4-$flow. In both cases,
the theory required the development of a systematic formulation of covariant
gyrokinetic theory for the appropriate Lagrangian variational description of
single-particle dynamics in relativistic plasma regimes, permitting one to
identify a non-perturbative representation of the particle magnetic moment,
which was shown to be conserved even when global space-time symmetries may
be absent.

It is worth mentioning that additional contributions of KDF and fluid
anisotropies in plasmas can originate also from the conservation of particle
canonical momentum and by the occurrence of epicyclic motion. The first of
these two mechanisms arises primarily in both axisymmetric systems
characterized by strong rotation phenomena (velocity shear) or supersonic
flows \cite{Cr2013c}, as well as in the framework of energy-independent
kinetic\ equilibria \cite{PoP2014-1,PRE-new}. The second one instead affects
particles in axisymmetric systems and simultaneously exhibiting epicyclic
motion around minima of effective potential \cite{Cr2013b,newa,newb} and
related magnetic string configurations \cite{St3,St4}.

The search of physical sources for phase-space anisotropies in the case of
neutral matter can only rely on the role of the gravitational field, since
mechanisms related to magnetic fields are necessarily ruled out. In this
reference, a kinetic theory for spherical stellar systems with spheroidal
velocity distributions was proposed in Ref.\cite{r85}. In addition, a study
of the statistical properties of collisionless non-relativistic DM systems
at equilibrium can be found in Ref.\cite{IJMPD}, which is based on a kinetic
theory developed in the framework of the Vlasov--Poisson description. It was
shown that structures of this type are generally characterized by
intrinsically non-Maxwellian KDFs and exhibit temperature anisotropy, to be
intended as an anisotropy in the directional particle velocity dispersions.
It was proved that this feature can arise at equilibrium due to
specifically-kinetic effects associated with nonuniform gravitational fields
together with phase-space conservation laws.

The latter ones include in particular the mechanism based on the
conservation of angular momentum carried by the equilibrium KDF. This
feature is well-known to apply also in curved space-time and has been
studied in the case of $N-$body systems belonging to spherically-symmetric
equilibrium configurations, like collisionless stellar clusters (see Ref.%
\cite{gravi}). As a consequence, in both relativistic and non-relativistic
regimes, the resulting effect of such a functional dependence for the KDF is
the generation of an anisotropy between radial and tangential pressures,
when referred to an inertial reference frame in terms of spherical
coordinates.

These results demonstrate that non-isotropic velocity dependences carried by
the KDF as well as corresponding non-isotropic pressure tensors are expected
to naturally arise in collisionless $N-$body systems, while their origin is
made manifest and their physical properties are understood when a
statistical treatment based on kinetic theory is adopted. In particular, the
mechanisms relying on the existence of configuration-space symmetries can
provide a transversal framework and are expected to become relevant
especially for uncharged systems of the type mentioned above. The problem
posed then is whether other effects of similar type of those listed above
can be identified yielding phase-space anisotropies, to be possibly
characteristic only of relativistic systems in curved space-time, and such
that they realize additional physical sources for the generation of
corresponding non-isotropic kinetic and fluid equilibria in collisionless $N-
$body systems composed by uncharged matter.

\section{2 - Goals of the study}

The previous considerations on the physical origin of non-isotropic kinetic
and fluid equilibrium solutions provide the premises for the present work.
The target here is to point out the existence of a particular source of
anisotropy in relativistic collisionless $N-$body systems at kinetic
equilibrium, which is carried by the equilibrium KDF and can manifest itself
through its continuum fluid moments, in particular the stress-energy tensor.
The new physical mechanism is shown to exhibit the following notable
features:

A)\ It relies on the existence of the so-called Carter constant for the
single-particle dynamics around black-holes. As such, it applies to curved
space-times in which self-gravity effects of the $N-$body system can be
ignored (test matter), while the background metric tensor describes a
stationary solution of the Einstein equations characterized by either axial
symmetry, like the Kerr and Kerr-Newman solutions, or spherical symmetry as
the limiting case, i.e. the Schwarzschild and Reissner-Nordstr\"{o}m
space-times.

B) It leads to equilibrium KDFs that differ from the isotropic relativistic
Maxwellian distribution, but still allowing for the existence of
Gaussian-like distributions. The KDFs determined in this way are
intrinsically non-isotropic as far as their dependence on particle $4-$%
velocity components is concerned.

C) It gives rise to non-isotropic equilibrium stress-energy tensors, so that
when evaluated in a prescribed reference frame it is responsible for the
occurrence of a temperature anisotropy effect among all the spatial
directions. The outcome in such a case is formally analogous to the pressure
anisotropy mechanism produced in plasmas by the conservation of particle
magnetic moment.

D)\ It applies to both spherically-symmetric as well as axially-symmetric
systems. This generalizes both the analogous non-relativistic mechanism
proposed in Ref.\cite{IJMPD} for axially-symmetric geometries as well as the
mechanisms previously pointed out in spherically-symmetric configurations.

In detail, the goals of the paper are as follows:

1)\ To properly define the concepts of isotropic and non-isotropic kinetic
solutions, in reference both to the form of the KDF and of its corresponding
continuum tensorial fluid fields.

2)\ To determine the functional form of the solution of the equilibrium KDF
that describes collisionless $N-$body systems of neutral particles in terms
of the complete set of integrals of motion. For an illustration of the
theory, the case of background Kerr space-time is considered, implying
conservation laws for the single-particle energy, the angular momentum and
the Carter constant. The technique based on the method of invariants\ (see
Refs.\cite{Cr2010,Cr2011} and \cite{PoP2014-1,PoP2014-2} for its
relativistic generalization) is adopted, which requires the same KDF to be
expressed as a function of the integrals of motion only.

3)\ To propose an explicit representation of the equilibrium KDF in terms of
a strictly-positive smooth Gaussian-like distribution and to determine the
physical sources of phase-space anisotropy.

4)\ To determine an analytical approximation of the exact solution which
applies under suitable ordering conditions, to be referred to as
weak-anisotropy regime. The target of this analysis is to gain insight into
the physical properties of the new Carter-constant anisotropy mechanism.

5)\ To evaluate the stress-energy tensor associated with the equilibrium KDF
and to show that this is non-isotropic, implying the occurrence of a
temperature anisotropy effect in the collisionless $N-$body system. The
analysis is performed invoking the weak-anisotropy regime for the KDF. It is
pointed out that, for the type of solution considered here, this mechanism
arises uniquely due to the existence of the Carter constant for the particle
geodesic motion and its inclusion in the equilibrium KDF. Hence, this
represents an intrinsically-relativistic effect that can be only inspected
in the framework of a covariant statistical Vlasov theory and which applies
to axially-symmetric space-time solutions as well as to
spherically-symmetric geometries as their limiting case.

\section{3 - Isotropic and non-isotropic kinetic solutions}

In this section we introduce the concepts of isotropic and non-isotropic
kinetic solutions for collisionless $N-$body systems in the framework of the
covariant Vlasov theory. These refer both to the properties of the KDF as
well as to the form of the corresponding fluid moments, in particular the
stress-energy tensor $T^{\mu \nu }\left( r^{\alpha }\right) $.

In order to address the issue and without loss of generality, we consider
the case of a single-species $N-$body system composed of neutral particles
having rest mass $m$, so that the subscript $\mathrm{j}$ of the species
index will be omitted in the following. We also treat the case of particles
having unitary rest mass, so that hereon $m=1$. The generalization to a
multi-species configuration becomes straightforward in the framework of
collisionless statistical dynamics, once the single-species solution is
obtained. For this reason, in reference to the astrophysical context, the
model proposed here is susceptible of applications either to DM halos or to
stellar clusters. It is assumed that the same $N-$body system is orbiting
around a compact object of rest mass $M$ (in the case of a stellar system
this corresponds to assume a supermassive black-hole of $M\gtrsim
10^{6}M_{\odot }$).

The generic particle has a $4-$position $r^{\mu }$ and a $4-$velocity $%
u^{\mu }\equiv \frac{dr^{\mu }}{ds}$, with $s$ being the corresponding
world-line proper time. By construction the line element $ds$ satisfies the
identity%
\begin{equation}
ds^{2}=g_{\mu \nu }\left( r^{\alpha }\right) dr^{\mu }dr^{\nu },  \label{ds}
\end{equation}%
where $g_{\mu \nu }\left( r^{\alpha }\right) \in \mathcal{R}^{4}$ is the
position-dependent metric tensor which, according to Ref.\cite{wald}, is
taken to have signature $g_{\mu \nu }=\left( -,+,+,+\right) $, while $%
\mathcal{R}^{4}$ is the $4-$dimensional smooth Lorentzian manifold. The
explicit representation of the metric tensor will be given in the next
section. Eq.(\ref{ds}) implies the mass-shell constraint for the $4-$%
velocity $u^{\mu }$, namely%
\begin{equation}
g_{\mu \nu }\left( r^{\alpha }\right) u^{\mu }u^{\nu }=1.
\label{mass-shell1}
\end{equation}%
This must be intended as identically satisfied along the particle geodesic
trajectory determined by the equation of motion%
\begin{equation}
\frac{D}{Ds}u^{\mu }\left( s\right) =0,
\end{equation}%
where $\frac{D}{Ds}$ denotes the covariant derivative and the proper-time
parametrization is adopted for\ $u^{\mu }$, consistent with Eq.(\ref{ds}).
We then denote the particle Lagrangian state $\mathbf{x}\left( s\right) $
with $\mathbf{x}\left( s\right) \equiv \left( r^{\mu }\left( s\right)
,u^{\mu }\left( s\right) \equiv \frac{dr^{\mu }\left( s\right) }{ds}\right) $%
.

For the covariant representation of the particle dynamics and the subsequent
investigation of kinetic equilibria we adopt a formalism based on the
introduction of a tetrad of unit $4-$vectors, in analogy with the approach
followed in the case of relativistic plasmas in Refs.\cite%
{PoP2014-2,Bek1,Bek2}. Thus, we introduce the set of orthogonal unit $4-$%
vectors $\left( a^{\mu },b^{\mu },c^{\mu },d^{\mu }\right) $ which are
defined by the coordinate system, where $a^{\mu }$ and $\left( b^{\mu
},c^{\mu },d^{\mu }\right) $ are respectively time-like and space-like. In
terms of the unit $4-$vectors representation the particle $4-$velocity can
therefore be decomposed as%
\begin{equation}
u^{\mu }\equiv u_{0}a^{\mu }+u_{1}b^{\mu }+u_{2}c^{\mu }+u_{3}d^{\mu },
\label{4-veeldeco}
\end{equation}%
to be denoted as tetrad representation, where $\left(
u_{0},u_{1},u_{2},u_{3}\right) $ are the corresponding components and are
separately $4-$scalars. Then, we recall that for single-particle dynamics
the $4-$dimensional velocity space $\mathcal{V}^{4}$ is defined as the
tangent bundle of the configuration-space manifold $\mathcal{R}^{4}$ whose $%
4-$vectors are subject to the mass-shell constraint (\ref{mass-shell1}).
Thus, the unconstrained 8-dimensional phase-space $\Omega $ is defined as $%
\Omega =\mathcal{R}^{4}\times \mathcal{V}^{4}$. By construction, the tetrad
representation provides a local mutual relationship among the components of
the $4-$velocity (\ref{4-veeldeco}). In particular, this yields the
representation%
\begin{equation}
u_{0}=\sqrt{u_{1}^{2}+u_{2}^{2}+u_{3}^{2}-1},  \label{u-zero}
\end{equation}%
so that the time-component of the $4-$velocity depends on the squared
space-components. Notice that this equation is related to the mass-shell
condition (\ref{mass-shell1}).

Based on these premises, we can proceed with the appropriate definitions of
isotropic and non-isotropic kinetic solutions. We consider first the
distribution function. In the following a generic relativistic KDF $f$ will
be said to be isotropic on the velocity space $\mathcal{V}^{4}$ if it
carries even powers of all the $4-$velocity components $\left(
u_{1},u_{2},u_{3}\right) $ and this functional dependence is isotropic. A
particular case of isotropic dependence is through the velocity component $%
u_{0}$, since the relationship (\ref{u-zero}) always applies locally. Hence,
a KDF of the type $f=f\left( u_{0}\right) $ represents an isotropic
distribution. A notable example of isotropic KDF of this form is the
relativistic Maxwellian function $f_{M}$. An exhaustive treatment of the
Maxwellian solution can be found in Ref.\cite{degroot}. In contrast, a KDF $%
f $ will be said to be non-isotropic on $\mathcal{V}^{4}$ if it exhibits a
non-isotropic dependence on the even powers of the $4-$velocity components $%
\left( u_{1},u_{2},u_{3}\right) $. Examples of non-isotropic KDF of this
type for relativistic plasma kinetic equilibria can be found in Refs.\cite%
{PoP2014-1,PoP2014-2}. It is clear from these considerations that
non-isotropic distribution functions necessarily differ from local
Maxwellian distributions, while from a statistical point of view such
deviations are due to velocity-space, or more generally phase-space kinetic
effects arising from single-particle dynamics.

As mentioned above, in kinetic theory the observable fluid fields of a given
$N-$body system can be determined \textquotedblleft a
posteriori\textquotedblright\ once the kinetic solution for the KDF and its
dynamical behavior are known. This is accomplished by defining the same
fluid fields as appropriate velocity integrals of the KDF weighted by
suitable phase-space dependent weight functions. This approach marks the
strict connection between kinetic and fluid approaches and gives physical
meaning to both the corresponding mathematical solutions. In the present
work we are particularly interested in considering the stress-energy tensor $%
T^{\mu \nu }\left( r^{\alpha }\right) $ for the continuum fluid system. This
is defined by the $4-$velocity integral%
\begin{equation}
T^{\mu \nu }\left( r^{\alpha }\right) =2\int_{\mathcal{V}^{4}}\sqrt{-g}%
d^{4}u\Theta \left( u^{0}\right) \delta \left( u^{\mu }u_{\mu }-1\right)
u^{\mu }u^{\nu }f,  \label{tmunu}
\end{equation}%
where the Dirac-delta takes into account the kinematic constraint (\ref%
{mass-shell1}) for the $4-$velocity when performing the integration, $\Theta
$ denotes the Theta-function selecting the root of $u^{0}$, while $\sqrt{-g}$
is the square-root of the determinant of the background metric tensor.
Invoking the tetrad representation for the $4-$velocity (\ref{4-veeldeco}),
the integral (\ref{tmunu}) can be reduced to%
\begin{equation}
T^{\mu \nu }\left( r^{\alpha }\right) =\int \frac{\sqrt{-g}d^{3}u}{\sqrt{%
u_{1}^{2}+u_{2}^{2}+u_{3}^{2}-1}}u^{\mu }u^{\nu }f,  \label{tmunu-bis}
\end{equation}%
which is now defined over the $3-$dimensional tangent space in which the
component $u_{0}$ of the $4-$velocity must be intended as dependent on the
other components according to Eq.(\ref{u-zero}). In the following, a
stress-energy tensor $T^{\mu \nu }\left( r^{\alpha }\right) $ is said to be
isotropic if it is defined in terms of an isotropic KDF $f$.
Correspondingly, the same tensor field will be referred to as non-isotropic
stress-energy tensor if the KDF corresponding to the collisionless $N-$body
system and entering Eq.(\ref{tmunu-bis}) is a non-isotropic distribution, in
the sense defined above. Thus, an example of isotropic stress-energy tensor
is the one associated with a relativistic Maxwellian KDF $f_{M}$ which
describes ideal fluids, to be denoted as $T_{M}^{\mu \nu }\left( r^{\alpha
}\right) $. This takes the customary tensorial form%
\begin{equation}
T_{M}^{\mu \nu }\left( r^{\alpha }\right) =neU^{\mu }U^{\nu }-p\Delta ^{\mu
\nu },  \label{T-max}
\end{equation}%
where $n$ is the particle number density and $e$ is the energy per particle,
so that $ne$ is the fluid energy density, $U^{\mu }$ is the fluid $4-$%
velocity, $p$ is the scalar isotropic pressure and $\Delta ^{\mu \nu }$ is
the so-called projector operator $\Delta ^{\mu \nu }=g^{\mu \nu }-U^{\mu
}U^{\nu }$. Again, we refer to Ref.\cite{degroot} for a comprehensive
treatment of Maxwellian fluid fields and their explicit analytical
calculation.

\section{4 - Carter constant in Kerr metric}

In this section we present the set of appropriate integrals of motion
required for the treatment of collisionless equilibria realized by
collisionless matter orbiting around a compact object.

To introduce the issue, we preliminary notice that the complete solution of
the metric tensor $g_{\mu \nu }\left( r^{\alpha }\right) $ can in principle
be determined by solving the coupled set of Einstein-Vlasov equations, for a
given distribution of source matter, consisting of the compact object and
the collisionless neutral matter, the latter being identified by the
corresponding KDF. However, in the present work we consider an $N-$body
system whose gravitational contribution to $g_{\mu \nu }\left( r^{\alpha
}\right) $ is negligible with respect to the background\ one due to the
compact object. In such a case, in the first approximation the Einstein and
Vlasov equations decouple, so that the metric tensor can be solved and
prescribed independently, and the $N-$body system is treated as a system
testing the influence of the given black-hole space-time.

We consider space-time solutions for the background metric tensor which are
characterized by symmetry properties, to be expressed in terms of Killing
vector fields \cite{wald}. Configurations admitting both Killing vectors and
Killing tensors are of interest here. A generic Killing vector $C_{\mu }$
satisfies the Killing equation $C_{\mu ;\nu }+C_{\nu ;\mu }=0$, where ";"
denotes covariant differentiation in standard notation, and is associated
with the symmetry of space-time with respect to a cyclic (i.e., ignorable)
coordinate. To each Killing vector $C^{\mu }$ a conserved quantity of the
geodesic motion corresponds, which is linear in $u^{\mu }$ and is given by
the $4-$scalar $C\equiv C^{\mu }u_{\mu }$. Such a type of conservation law
therefore arises naturally whenever the metric tensor admits cyclic
coordinates. On the other hand, Killing tensors are associated with
so-called hidden symmetries, and their existence must be ascertained for a
given metric solution. A symmetric second-rank Killing tensor $K_{\mu \nu }$
satisfies the Killing equation $K_{\mu \nu ;\alpha }+K_{\mu \alpha ;\nu }=0$
and generates a quadratic integral of motion expressed by the 4-scalar $%
K\equiv K_{\mu \nu }u^{\mu }u^{\nu }$ (as in the case of the Carter
constant). This type of dependence on the particle $4-$velocity carried by $K
$ is of crucial importance for the occurrence of the temperature-anisotropy
mechanism displayed in this work.

For an illustration of the theoretical model, we consider here an uncharged
rotating black hole of rest mass $M$ and intrinsic angular momentum $S$
described by the Kerr metric. Adopting Boyer-Lindquist coordinates $\left(
t,\phi ,r,\theta \right) $ and the geometrical system of units ($c=G=1$),
the corresponding metric is expressed in terms of the line element (\ref{ds}%
) as%
\begin{eqnarray}
ds^{2} &=&-\left[ \frac{\Delta -a^{2}\sin ^{2}\theta }{\Sigma }\right]
dt^{2}-\frac{4Mar\sin ^{2}\theta }{\Sigma }dtd\phi  \notag \\
&&+\left[ \frac{\left( r^{2}+a^{2}\right) ^{2}-\Delta a^{2}\sin ^{2}\theta }{%
\Sigma }\right] \sin ^{2}\theta d\phi ^{2}+\frac{\Sigma }{\Delta }%
dr^{2}+\Sigma d\theta ^{2},  \label{kerr}
\end{eqnarray}%
where%
\begin{eqnarray}
\Sigma &\equiv &r^{2}+a^{2}\cos ^{2}\theta , \\
\Delta &\equiv &r^{2}+a^{2}-2Mr,
\end{eqnarray}%
with $a\equiv S/M$ being the angular momentum per unit mass.
Correspondingly, when adopting Boyer-Lindquist coordinates we shall denote
the velocity components $\left( u_{0},u_{1},u_{2},u_{3}\right) $ in Eq.(\ref%
{4-veeldeco}) respectively as $\left( u_{0},u_{1},u_{2},u_{3}\right) \equiv
\left( u_{t},u_{\phi },u_{r},u_{\theta }\right) $.

Thanks to the stationarity and axisymmetry assumptions underlying the Kerr
solution, the coordinates $t$ and $\phi $ are ignorable. This implies that
there are two Killing vectors, to be denoted as $\xi ^{\alpha }=\left(
\partial /\partial t\right) ^{\alpha }$ and $\zeta ^{\alpha }=\left(
\partial /\partial \phi \right) ^{\alpha }$, which determine corresponding
integrals of motion to be identified respectively with the total particle
energy $E$ and angular momentum $L$. Invoking Eq.(\ref{kerr}), the latter
are given (per unit rest mass for particle geodesic) by%
\begin{equation}
E\equiv -\xi ^{\mu }u_{\mu }=\left[ 1-\frac{2Mr}{\Sigma }\right] \overset{%
\cdot }{t}+\frac{2Mar\sin ^{2}\theta }{\Sigma }\overset{\cdot }{\phi },
\label{E}
\end{equation}%
\begin{equation}
L\equiv \zeta ^{\mu }u_{\mu }=-\frac{2Mar\sin ^{2}\theta }{\Sigma }\overset{%
\cdot }{t}+\frac{\left( r^{2}+a^{2}\right) ^{2}-\Delta a^{2}\sin ^{2}\theta
}{\Sigma }\sin ^{2}\theta \overset{\cdot }{\phi },  \label{L}
\end{equation}%
where $\overset{\cdot }{t}=\frac{dt}{ds}\equiv u_{t}$ and $\overset{\cdot }{%
\phi }=\frac{d\phi }{ds}\equiv u_{\phi }$. In addition, the Kerr metric
admits a non-trivial Killing tensor, whose physical meaning is related to
the angular momentum of the field source. This generates the following
integral of the geodesic motion:%
\begin{equation}
K\equiv \mathcal{Q}+\left( L-aE\right) ^{2},  \label{K}
\end{equation}%
where $\mathcal{Q}$ denotes the \textit{Carter constant }\cite{Carter}%
\begin{equation}
\mathcal{Q=}p_{\theta }^{2}+\cos ^{2}\theta \left[ a^{2}\left(
1-E^{2}\right) +\left( \frac{L}{\sin \theta }\right) ^{2}\right] ,
\end{equation}%
and here $p_{\theta }=\Sigma \overset{\cdot }{\theta }$, with $\overset{%
\cdot }{\theta }=\frac{d\theta }{ds}\equiv u_{\theta }$. The scalar $K$ can
be shown to be always non-negative \cite{gravi}. For the following
developments, it is useful to equivalently represent the invariant $K$ as a
polynomial of the particle velocity components as%
\begin{equation}
K=\Sigma ^{2}u_{\theta }^{2}+A_{1}u_{t}^{2}+A_{2}u_{\phi
}^{2}+A_{3}u_{t}u_{\phi }+A_{4},  \label{k-poli}
\end{equation}%
where $A_{1}$, $A_{2}$, $A_{3}$ and $A_{4}$ are configuration-space
functions, whose expression follows from Eqs.(\ref{E})-(\ref{K}) and are
given respectively by%
\begin{eqnarray}
A_{1} &\equiv &a^{2}\sin ^{2}\theta , \\
A_{2} &\equiv &\left( r^{2}+a^{2}\right) ^{2}\sin ^{2}\theta , \\
A_{3} &\equiv &-2a\left( r^{2}+a^{2}\right) \sin ^{2}\theta , \\
A_{4} &\equiv &a^{2}\cos ^{2}\theta .
\end{eqnarray}%
Finally, the last conservation law is represented by Eq.(\ref{mass-shell1}).

In summary, for the case considered here of test fluid composed by
collisionless neutral matter orbiting around a Kerr black-hole, there are
three integrals of motion which constraint single-particle dynamics and
which are associated with corresponding space-time symmetries. These are
identified with the particle total energy $E$, the particle angular momentum
$L$ and the Carter constant $Q$ (or equivalently the combination $K$ in Eq.(%
\ref{K})). This set of invariants together with Eq.(\ref{mass-shell1})
represent the basis for the construction of kinetic equilibria developed in
Section 6.

\section{5 - Adiabatic invariants}

For the uncharged single-particle dynamics in the Kerr geometry the set $%
I\left( \mathbf{x}\left( s\right) \right) =\left( E,L,K\right) $ identifies
three exact particle invariants or integrals of motion, namely dynamical
quantities which are exactly conserved along the particle geodesic motion.
In particular, the set $I\left( \mathbf{x}\left( s\right) \right) $ together
with the mass-shell constraint (\ref{mass-shell1}) make the geodesic motion
integrable. As indicated, these invariants generally depend on the particle
state, namely on configuration-space and velocity-space variables. Thus, the
scalars $I\left( \mathbf{x}\left( s\right) \right) $ necessarily satisfy the
equation%
\begin{equation}
\frac{DI\left( \mathbf{x}\left( s\right) \right) }{Ds}=0.
\label{exact-invariants}
\end{equation}%
More generally, one can envisage a configuration in which one or more
invariants of the set $I\left( \mathbf{x}\left( s\right) \right) $ may be
conserved in asymptotic way, namely such that in place of Eq.(\ref%
{exact-invariants}), the equation%
\begin{equation}
\frac{DI\left( \mathbf{x}\left( s\right) \right) }{Ds}=0\left[ 1+O\left(
\varepsilon ^{h}\right) \right]   \label{adiabatic-invariant}
\end{equation}%
holds. Here $\varepsilon $ identifies a suitable dimensionless invariant
parameter of the system, while $h\geq 1$. In detail, the frame-invariant
parameter $\varepsilon $ is defined here as the ratio%
\begin{equation}
\varepsilon \equiv \frac{r_{L}}{R_{L}}\ll 1,  \label{epsi}
\end{equation}%
to be considered as an infinitesimal. Both $r_{L}$ and $R_{L}$ identify $4-$%
scalars which, in the present context, can be conveniently introduced
respectively as%
\begin{equation}
\frac{1}{r_{L}^{2}}\equiv \frac{1}{\lambda _{1}^{2}}\partial _{\mu }\lambda
_{1}\partial ^{\mu }\lambda _{1},
\end{equation}%
\begin{equation}
\frac{1}{R_{L}^{2}}\equiv \frac{1}{\lambda _{2}^{2}}\partial _{\mu }\lambda
_{2}\partial ^{\mu }\lambda _{2},
\end{equation}%
where $\lambda _{1}$ is associated with a physical observable of the system,
like the $4-$scalar temperature (see below), while $\lambda _{2}\equiv
R^{\alpha \beta \gamma \delta }R_{\alpha \beta \gamma \delta }$ is the
Kretschmann invariant constructed in terms of the Riemann tensor associated
with the background metric tensor. From the physical point of view, the
definition (\ref{epsi}) means that the physical properties of the $N-$body
system vary on smaller scales than the local characteristic scale $R_{L}$
associated with the background curved space-time. This notion is also
consistent with the adoption of a kinetic theory for a test fluid system.

Invariants which satisfy Eq.(\ref{adiabatic-invariant}) are referred to as
adiabatic invariants: they are dynamical quantities which are conserved
along geodesic motion in asymptotic sense, namely up to $O\left( \varepsilon
^{h}\right) $. Adiabatic invariants of the type considered here arise when
the configuration-space symmetries of the system are not exact, but can only
be defined in asymptotic way. The introduction of the notion of adiabatic
invariants is convenient in order to investigate more realistic physical
systems, since in nature conservation laws should be intended as asymptotic.
Thus, for example, $E$, $L$ and $K$ might be adiabatic invariants if
respectively the system is not exactly stationary or axisymmetric (with
respect to the coordinates $t$ and $\varphi $) or the metric tensor weakly
differs from the exact Kerr solution. Deviations of this type are expected
to contribute once the coupled set of Einstein-Vlasov equations is solved to
take into account the $N-$body self-gravity contribution to the system
metric tensor. The notion of adiabatic invariants permits therefore in
principle the proper perturbative treatment of these effects in the
framework of an asymptotic kinetic theory (see discussion reported below).

Based on these premises, in the following we shall investigate the role of
the invariants $I\left( \mathbf{x}\left( s\right) \right) $ as far as the
statistical treatment of the collisionless $N-$body system and the
generation of non-isotropic pressure tensors are concerned.

\section{6 - Relativistic non-isotropic kinetic equilibria}

In this section the realization of relativistic kinetic equilibria for
collisionless $N-$body systems of neutral matter in Kerr space-time is
considered. Following the approach developed in Refs.\cite%
{Cr2010,Cr2011,Cr2011a} for non-relativistic systems and in Refs.\cite%
{PoP2014-1,PoP2014-2}\textbf{\ }for plasmas in curved space-time, the method
of invariants is implemented. This consists in expressing the equilibrium
KDF in terms of the single-particle exact or adiabatic invariants. In fact,
since Eqs.(\ref{exact-invariants}) and (\ref{adiabatic-invariant}) apply
respectively for integrals of motion and adiabatic invariants, necessarily
any function which depends only on these quantities must be similarly
conserved along particle trajectory.

In view of the symmetry properties of the configuration considered here and
the assumptions introduced in the previous section, the invariants are
identified with the set $I\left( \mathbf{x}\left( s\right) \right) =(E,L,K)$
defined respectively by Eqs.(\ref{E}), (\ref{L}) and (\ref{K}). Therefore
one can always represent the equilibrium KDF in the form $f=f_{\ast }$, with%
\begin{equation}
f_{\ast }=f_{\ast }\left( \left( E,L,K\right) ,\Lambda _{\ast }\right)
\label{f-star}
\end{equation}%
being a smooth strictly-positive function of the particle invariants only,
which is summable in velocity-space. Concerning the notation, in Eq.(\ref%
{f-star})\ the set $\left( E,L,K\right) $ denotes explicit functional
dependences\ carried by the KDF on the invariants, while $\Lambda _{\ast }$
denotes the so-called structure functions \cite{Cr2011}, namely functions
suitably related to the observable velocity moments of the KDF (i.e., the
physical fluid fields of the system) that have to be defined below. By
construction, the structure functions carry implicit functional dependences
on the same set of invariants. Their general representation is therefore of
the form%
\begin{equation}
\Lambda _{\ast }=\Lambda _{\ast }\left( E,L,K,\varepsilon ^{b}r^{\mu
}\right) ,  \label{str-funcdep}
\end{equation}%
where $\varepsilon ^{b}r^{\mu }$ denotes a possible slow configuration-space
dependence and $b\geq 1$ is a suitable real number. We notice that such a
slow dependence, if included in the set $\Lambda _{\ast }$, necessarily
implies that the equilibrium KDF\ $f_{\ast }$ is an adiabatic invariant. As
a particular realization, Eq.(\ref{str-funcdep}) includes the case in which $%
\Lambda _{\ast }$ are identically constant, namely $\Lambda _{\ast }=const.$

From the definition of the set $I\left( \mathbf{x}\left( s\right) \right) $
and the type of functional dependences allowed for the equilibrium solution (%
\ref{f-star}) and the structure functions (\ref{str-funcdep}), it follows
that the KDF $f_{\ast }$ depends generally on the single-particle state $%
\mathbf{x}\left( s\right) $ only via the same invariants $I\left( \mathbf{x}%
\left( s\right) \right) $ and $\Lambda _{\ast }$, so that $f_{\ast }\left(
I\left( \mathbf{x}\left( s\right) \right) ,\Lambda _{\ast }\right) \equiv
\widehat{f}_{\ast }\left( \mathbf{x}\left( s\right) \right) $. In the
framework of the perturbative approach introduced above, a KDF of the form (%
\ref{f-star}) satisfies therefore the Vlasov equation in Lagrangian form%
\begin{equation}
\frac{d}{ds}\widehat{f}_{\ast }\left( \mathbf{x}\left( s\right) \right) =0%
\left[ 1+O\left( \varepsilon ^{d}\right) \right] ,
\end{equation}%
where\ the exponent $d$ determines the order of adiabatic invariant of the
equilibrium KDF $f_{\ast }$. More precisely, in view of the formalism
introduced above, the exponent $d$ must be chosen as the minimum between the
indices $h$ associated with the possible adiabatic character of the
invariants themselves (see Eq.(\ref{adiabatic-invariant})) and $b$ (see Eq.(%
\ref{str-funcdep})). As a consequence, by construction the KDF $f_{\ast }$
is generally an adiabatic invariant of $O\left( \varepsilon ^{d}\right) $.

As it is well-known, in the framework of the Vlasov theory the choice of the
form of the equilibrium KDF $f_{\ast }$ is not unique. Basic requirements
are that $f_{\ast }$ must be a strictly-positive scalar function which is
summable over the velocity space, to warrant the existence of related fluid
fields, together with the prescription of its admissible functional
dependence, which is taken here in accordance with Eq.(\ref{f-star}). Thus,
following previous analogous treatments carried out in the case of
relativistic collisionless plasmas, it is possible to introduce an explicit
representation for the same KDF satisfying these requirements. This is
expressed in terms of a Gaussian-like distribution as%
\begin{equation}
f_{\ast }=\beta _{\ast }e^{-E\gamma _{\ast }-L\omega _{\ast }-K\alpha _{\ast
}}.  \label{equil-gaussian}
\end{equation}%
The notation is as follows. First, $E$, $L$ and $K$ are the $4-$scalar
integrals of motion associated with the existence of corresponding Killing
vectors and tensors for the Kerr metric, and which are represented in
Boyer-Lindquist coordinates by Eqs.(\ref{E})-(\ref{K}). Second, the ensemble
$\Lambda _{\ast }=\left( \beta _{\ast },\gamma _{\ast },\omega _{\ast
},\alpha _{\ast }\right) $ identifies the set of structure functions, which
are assumed to be $4-$scalars by construction and whose general functional
dependence is according to Eq.(\ref{str-funcdep}). These functions are
related to the system physical observables. In particular, $\beta _{\ast }$
is associated with the system number density measured in the fluid comoving
frame, $\gamma _{\ast }$ determines the system isotropic temperature, while $%
\omega _{\ast }$ enters the definition of the fluid angular frequency along
the direction $\phi $ of spatial symmetry, when measured by an inertial
observer. Finally, $\alpha _{\ast }$ is related to the anisotropy character
of the solution associated with $K$ and is related to the system temperature
(and pressure) anisotropy, as expressed by the corresponding fluid
stress-energy tensor.

The three functions $\left( \beta _{\ast },\gamma _{\ast },\omega _{\ast
}\right) $ are equilibrium fields which generally arise in all Maxwellian
distributions describing collisionless systems of neutral and/or charged
matter (i.e., plasmas) in stationary and axisymmetric configurations, which
exhibit azimuthal flow in the $\phi $ direction and whose KDF is allowed to
depend on the conserved single-particle energy \cite%
{Cr2010,Cr2012,Cr2011,Cr2011a}. In fact, as is well-known, any isotropic
Maxwellian KDF carries the three fluid fields of the fluid system identified
with the number density, flow velocity and isotropic temperature. In
addition to this, in the present case the function $\alpha _{\ast }$
determines a phase-space anisotropy contribution carried by the constant $K$%
, so that it generates a deviation from the isotropic case. The role of the
invariant $K$ in the Gaussian-like KDF $f_{\ast }$ in Eq.(\ref%
{equil-gaussian}) is analogous to that played by the particle magnetic
moment $m^{\prime }$ in relativistic plasmas being responsible for the
occurrence of a temperature anisotropy effect in those systems \cite%
{Cr2010,Cr2012,Cr2011,Cr2011a}.

From the prescription (\ref{f-star}) we can therefore say that the same
equilibrium KDF is non-isotropic. In summary, according to the definition
given in Section 3, the following two different sources of phase-space
anisotropy can be identified in the equilibrium solution (\ref%
{equil-gaussian}):

1)\ The first one is due to the implicit functional dependences contained in
the structure functions $\Lambda _{\ast }$ according to Eq.(\ref{str-funcdep}%
). This type of anisotropy contribution has been widely investigated in the
case of plasmas for different physical regimes and asymptotic orderings (see
Refs.\cite{Cr2010,Cr2011}). It is associated with the so-called diamagnetic,
finite Larmor-radius and energy effects described in the Introduction. In
particular, a notable example in which such an anisotropy effect enters the
solution is reported in Ref.\cite{Cr2013b}\textbf{\ }and concerns kinetic
equilibria in the presence of strong rotation phenomena, like supersonic
flows and/or shear flows.

2)\ The second one is due to the explicit functional dependence on the
invariant $K$. In fact the latter quantity is a non-isotropic function of
the squared of the velocity components $u_{\theta }$, $u_{t}$ and $u_{\phi }$%
, according to the representation given by Eq.(\ref{k-poli}).

\section{7 - Weak-anisotropy regime}

In this section we determine a suitable asymptotic approximation of the
exact equilibrium KDF (\ref{equil-gaussian}). The analysis is useful in
order to gain insight into the physical properties of the equilibria
considered here and to elucidate the related contribution of the
Carter-constant term. The approximation derived below is also propedeutic
for the subsequent study on the occurrence of temperature anisotropy in the
stress-energy tensor.

To start with, we introduce the following conditions on the equilibrium KDF:

1)\ The structure functions $\Lambda _{\ast }$ are assumed to be constant,
in the sense that%
\begin{equation}
\Lambda _{\ast }=\Lambda \left[ 1+O\left( \varepsilon ^{2}\right) \right] ,
\label{lambda-asym}
\end{equation}%
where $\Lambda =\left( \beta ,\gamma ,\omega ,\alpha \right) $ denotes the
set of constant values associated with\ $\Lambda _{\ast }$, so that here $%
\Lambda =const.$ In this case the possible functional dependences on the
adiabatic invariants carried by the same $\Lambda _{\ast }$ are considered
at most of 2nd-order in the parameter $\varepsilon $. This hypothesis allows
one to neglect the implicit phase-space contributions contained in $\Lambda
_{\ast }$, both for the kinetic solution as well as for the calculation of
the corresponding fluid fields. Indeed, these contributions are expected to
give rise to important kinetic effects analogous to the so-called
diamagnetic, finite Larmor-radius and energy-correction effects pointed out
in the case of plasmas, both in non-relativistic and relativistic regimes.
However, the treatment of this type of dependences would require the proper
set up of a perturbative theory analogous to the one developed in Refs.\cite%
{Cr2010,Cr2011,Cr2011a}. Instead, the advantage of the choice (\ref%
{lambda-asym}) on the structure functions is that it permits to single out
the role of the invariant $K$, which is unique of the equilibrium solution
of collisionless neutral matter in curved space-time, with respect to other
kinetic effects of the type mentioned above. It must be noticed that,
although the constant structure functions $\Lambda $ still remain related to
the observable fluid fields as described in the previous section, the
assumption (\ref{lambda-asym}) does not imply that the same fluid fields are
constant too. In fact, when Eq.(\ref{lambda-asym}) applies, the equilibrium
KDF still retains configuration-space dependences which are due to the
non-uniform background metric tensor and enter through the invariants $%
I\left( \mathbf{x}\left( s\right) \right) $, according to Eqs.(\ref{E})-(\ref%
{K}). These type of spatial dependences are then inherited by the continuum
fluid fields once integration of the KDF over the velocity space is
performed.

2)\ The asymptotic ordering%
\begin{equation}
K\alpha \sim O\left( \varepsilon \right)  \label{war}
\end{equation}%
applies in a suitable subset of phase-space. This condition can be realized
by properly setting the value of the constant $\alpha $. It amounts to
consider a configuration in which the contribution of the invariant $K$ to
the equilibrium solution is of higher-order with respect to the other
invariants. Because of this feature and since, as shown below, $K\alpha $ is
the term in the equilibrium KDF responsible for the occurrence of
temperature anisotropy, the ordering (\ref{war}) will be referred to here as
determining the \emph{weak-anisotropy regime}. As a consequence, in the
weak-anisotropy regime the anisotropy is treated as a perturbative effect.
It must be stressed nevertheless that such an ordering condition is adopted
here for an analytical illustration of the Carter anisotropy mechanism. Full
account of the effect without imposing any ordering condition can generally
be studied based on the exact solution (\ref{equil-gaussian})\ pointed out
in Section 6, together with Eq.(\ref{str-funcdep}).

When assumptions 1 and 2 apply, the equilibrium KDF $f_{\ast }$ in Eq.(\ref%
{equil-gaussian}) can be Taylor-expanded to first-order in $\varepsilon $ to
give%
\begin{equation}
f_{\ast }=f_{w}\left[ 1+O\left( \varepsilon ^{2}\right) \right] ,
\end{equation}%
where $f_{w}$ denotes the following distribution%
\begin{equation}
f_{w}=f_{M}\left( 1-K\alpha \right) ,  \label{war-kdf}
\end{equation}%
with $f_{M}$ being the Maxwellian function%
\begin{equation}
f_{M}\equiv \beta e^{-E\gamma -L\omega }.  \label{maxw}
\end{equation}

Hence, by construction in such a limit, $f_{w}$ represents an adiabatic
invariant of $O\left( \varepsilon \right) $. From the physical point of view
we notice that in the weak-anisotropy regime considered here the term $%
K\alpha $ is a polynomial of second degree on particle velocity which
provides a first-order correction to the leading-order isotropic Maxwellian
distribution $f_{M}$. As\ discussed below, this determines non-vanishing
contributions to the stress-energy tensor associated with $f_{\ast w}$,
giving rise to temperature anisotropy effect, to be considered weak in the
sense of Eq.(\ref{war}). Therefore, according to this representation the
anisotropic character of the equilibrium kinetic solution is uniquely and
completely associated with the polynomial $K\alpha $, so that the
Carter-constant term is responsible for the deviation from the customary
Maxwellian solution widely used in the literature. The conceptual difference
between isotropic Maxwellian solution $f_{M}$ and the non-isotropic one $%
f_{w}$\ is relevant and non-negligible, since the invariant $K$ is actually
a peculiar feature of a rotating black-hole, described by the Kerr solution.
Hence, the functional dependence prescribed for the equilibrium KDF $f_{\ast
}$ and its analytical treatment in terms of $f_{w}$ are relevant as they
include all the peculiarities of curved space-time solutions in General
Relativity and related particle dynamics.

\section{8 - Non-isotropic stress-energy tensor}

In this section we discuss the qualitative properties of the stress-energy
tensor corresponding to the anisotropic equilibrium KDF prescribed above.
For this purpose we adopt the representation given by Eq.(\ref{tmunu-bis}).
We first treat the case in which the KDF $f$ in the same equation coincides
with the exact equilibrium solution $f_{\ast }$. According to the
definitions given above, since the latter KDF is non-isotropic, it follows
that the corresponding tensor field $T^{\mu \nu }\left( r^{\alpha }\right) $
is non-isotropic too. The explicit evaluation of the components of such
tensor can be generally achieved by means of numerical integration, once the
preliminary prescription of the functional form of the structure functions
is done according to the precise physical system to be studied. From the
discussion reported in Section 5, one can see that in this case the
non-isotropic character of $T^{\mu \nu }\left( r^{\alpha }\right) $ is a
consequence of the combined action of the dependence of $f_{\ast }$ on the
invariant $K$ and the non-uniform phase-space profiles of the structure
functions. In turn, both these features are characteristic of the
relativistic equilibrium solution in the Kerr background metric tensor.

A deeper insight into the issue can be gained by considering the equilibrium
solution in the weak-anisotropy regime, since this permits to move one step
further with the analytical solution. In this limit, invoking Eq.(\ref%
{war-kdf}) and ignoring corrections of\ $O\left( \varepsilon ^{2}\right) $,
the asymptotic stress-energy tensor becomes%
\begin{equation}
T^{\mu \nu }\left( r^{\alpha }\right) =\int \frac{\sqrt{-g}d^{3}u}{\sqrt{%
u_{\phi }^{2}+u_{r}^{2}+u_{\theta }^{2}-1}}u^{\mu }u^{\nu }f_{M}\left(
1-K\alpha \right) ,  \label{T-weak}
\end{equation}%
where we have adopted Boyer-Lindquist coordinates and in the integral the
component $u_{t}$ is given by Eq.(\ref{u-zero}). Because of the series
representation of the equilibrium KDF, the tensor $T^{\mu \nu }\left(
r^{\alpha }\right) $ in Eq.(\ref{T-weak}) can be decomposed as%
\begin{equation}
T^{\mu \nu }\left( r^{\alpha }\right) =T_{M}^{\mu \nu }\left( r^{\alpha
}\right) +\Pi ^{\mu \nu }\left( r^{\alpha }\right) .
\end{equation}%
Here, $T_{M}^{\mu \nu }\left( r^{\alpha }\right) $ represents the
leading-order contribution associated with the Maxwellian KDF $f_{M}$. Its
expression therefore coincides with the one given by Eq.(\ref{T-max}). At
this order\ in this asymptotic evaluation, no contribution arises from the
invariant $K$, so that, in the weak-anisotropy regime, the leading-order
stress-energy tensor is isotropic. In contrast, the non-isotropic character
of $T^{\mu \nu }\left( r^{\alpha }\right) $ is carried by the $O\left(
\varepsilon \right) $ term $\Pi ^{\mu \nu }\left( r^{\alpha }\right) $,
which, from Eq.(\ref{T-weak}), is found to be%
\begin{equation}
\Pi ^{\mu \nu }\left( r^{\alpha }\right) =-\alpha \int \frac{\sqrt{-g}d^{3}u%
}{\sqrt{u_{\phi }^{2}+u_{r}^{2}+u_{\theta }^{2}-1}}u^{\mu }u^{\nu }Kf_{M}.
\end{equation}%
In this regime, the non-isotropic character of $\Pi ^{\mu \nu }\left(
r^{\alpha }\right) $ arises only due to the explicit dependence on the
invariant $K$. In fact, invoking the representation (\ref{k-poli}) given
above, one has that the integrand in the previous equation becomes a
polynomial function carrying a non-isotropic dependence on the particle
velocity components $\left( u_{\phi },u_{r},u_{\theta }\right) $. The same
polynomial is weighted on a Maxwellian KDF, a feature which warrants that it
is summable on the whole velocity space where the previous integral is
defined.

It is interesting to consider explicitly the contributions arising from $\Pi
^{\mu \nu }\left( r^{\alpha }\right) $, in such a way to prove its
non-isotropic character. We evaluate the same tensor adopting
Boyer-Lindquist coordinates and for this purpose we represent the particle $%
4-$velocity from Eq.(\ref{4-veeldeco}) in terms of the unit $4-$vectors $%
\left( a^{t},b^{\phi },c^{r},d^{\theta }\right) $ as%
\begin{equation}
u^{\mu }\equiv u_{t}a^{t}+u_{\phi }b^{\phi }+u_{r}c^{r}+u_{\theta }d^{\theta
}.
\end{equation}%
We remark that, once $u^{\mu }$ is represented in such a tetrad in terms of
the basis formed by $\left( a^{t},b^{\phi },c^{r},d^{\theta }\right) $, the
same $4-$vectors also identify the tensorial components of $T^{\mu \nu
}\left( r^{\alpha }\right) $, which are generally position-dependent.

Let us then consider the representation (\ref{k-poli}) for the invariant $K$%
. Manifestly, the integral on the configuration-space function $A_{4}$
yields simply a first-order isotropic contribution to the tensor $T_{M}^{\mu
\nu }\left( r^{\alpha }\right) $. The non-isotropic features arise therefore
from the terms proportional to $\Sigma ^{2}$, $A_{1}$, $A_{2}$ and $A_{3}$
in Eq.(\ref{k-poli}). We analyze first the diagonal components of $\Pi ^{\mu
\nu }\left( r^{\alpha }\right) $. In this reference, we notice that the term
proportional to $A_{3}$ does not contribute being odd in $u_{\phi }$. It
follows that the contribution $\Pi ^{tt}\left( r^{\alpha }\right) $ is given
by%
\begin{eqnarray}
\Pi ^{tt}\left( r^{\alpha }\right) &=&-\alpha a^{t}a^{t}\int \sqrt{-g}d^{3}u%
\sqrt{u_{\phi }^{2}+u_{r}^{2}+u_{\theta }^{2}-1}  \notag \\
&&\left[ \Sigma ^{2}u_{\theta }^{2}+A_{1}u_{t}^{2}+A_{2}u_{\phi }^{2}\right]
f_{M},  \label{pi-tt}
\end{eqnarray}%
while the component $\Pi ^{\phi \phi }\left( r^{\alpha }\right) $ is instead%
\begin{eqnarray}
\Pi ^{\phi \phi }\left( r^{\alpha }\right) &=&-\alpha b^{\phi }b^{\phi }\int
\frac{\sqrt{-g}d^{3}u}{\sqrt{u_{\phi }^{2}+u_{r}^{2}+u_{\theta }^{2}-1}}%
u_{\phi }^{2}  \notag \\
&&\left[ \Sigma ^{2}u_{\theta }^{2}+A_{1}u_{t}^{2}+A_{2}u_{\phi }^{2}\right]
f_{M}.  \label{pi-fifi}
\end{eqnarray}%
The remaining diagonal components $\Pi ^{rr}\left( r^{\alpha }\right) $ and $%
\Pi ^{\theta \theta }\left( r^{\alpha }\right) $ are formally analogous to $%
\Pi ^{\phi \phi }\left( r^{\alpha }\right) $ and can be obtained by
replacing in Eq.(\ref{pi-fifi}) the term $b^{\phi }b^{\phi }u_{\phi }^{2}$
respectively with $c^{r}c^{r}u_{r}^{2}$ and $d^{\theta }d^{\theta }u_{\theta
}^{2}$. It is immediate to conclude that all the diagonal terms of the
tensor are generally different from each other, the result of the
integration depending on the degree of the polynomial of the velocity
components over which the KDF $f_{M}$ is integrated.

Finally, contrary to the leading-order tensor $T_{M}^{\mu \nu }\left(
r^{\alpha }\right) $ which is purely diagonal, the first-order contribution $%
\Pi ^{\mu \nu }\left( r^{\alpha }\right) $ carries also non-vanishing
off-diagonal terms. These are the symmetric entries $\Pi ^{t\phi }\left(
r^{\alpha }\right) =\Pi ^{\phi t}\left( r^{\alpha }\right) $ which arise due
to the term $A_{3}u_{t}u_{\phi }$ in Eq.(\ref{k-poli}) and are given by%
\begin{equation}
\Pi ^{t\phi }\left( r^{\alpha }\right) =-\alpha a^{t}b^{\phi }A_{3}\int
\sqrt{-g}d^{3}u\sqrt{u_{\phi }^{2}+u_{r}^{2}+u_{\theta }^{2}-1}u_{\phi
}^{2}f_{M}.
\end{equation}%
The existence of the contribution $\Pi ^{t\phi }\left( r^{\alpha }\right) $
shows that the consistent treatment of collisionless equilibria and the
inclusion of the invariant $K$ in the equilibrium KDF determine a
non-trivial structure for the stress-energy tensor, in comparison with the
representation holding for isotropic Maxwellian distributions. In fact, a
first kind of anisotropy arises because the diagonal terms of the
stress-energy tensor differ from each other. Second, the same tensor is
generally non-diagonal when expressed in the Boyer-Lindquist coordinate
system.

To conclude the section, a comment is in order concerning the spatial
dependences in terms of $r^{\mu }$ arising in $T^{\mu \nu }\left( r^{\alpha
}\right) $. In fact, even in the case in which the structure functions
entering the equilibrium KDF are constant, like the weak-anisotropy regime,
non-trivial configuration-space dependences still arise due to the
non-homogeneous background Kerr space-time, as expressed by the metric
tensor represented here in Boyer-Lindquist coordinates.

\section{9 - Conclusions}

In this paper the discovery of an intrinsically-relativistic mechanism for
the generation of non-isotropic kinetic equilibria in collisionless $N-$body
systems of uncharged matter has been reported. The new effect relies on the
constraints placed by the single-particle conservation laws in
non-homogeneous curved space-times which admit the existence of Killing
tensors for the particle geodesic motion. To illustrate the issue, in the
present work the case of a collisionless system in a Kerr space-time has
been considered, assuming its self-gravity contribution to be negligible
with respect to the background one. The theory has been developed in the
framework of the covariant Vlasov statistical description, following recent
analogous developments carried out for relativistic astrophysical plasmas.
The construction of equilibrium kinetic distribution functions has relied on
the method of invariants, consisting in expressing them in terms of the
particle invariants only. For the case of interest, these were represented
by the particle energy, angular momentum and by the so-called Carter
constant. It has been pointed out that kinetic equilibria constructed in
this way necessarily correspond to continuum fluid equilibria, for which the
prescription of equilibrium closure conditions (e.g., the stress-energy
tensor) is consistently fulfilled.

In this regard, the precise meaning of isotropic and non-isotropic kinetic
and fluid equilibria in the context of a covariant statistical theory have
been introduced. Then, the general functional form of the equilibrium
distribution function has been determined together with an explicit
realization in terms of a Gaussian-like distribution. The physical
properties of these solutions have been investigated, with particular
emphasis on the meaning of both implicit and explicit phase-space functional
dependences expressed by the set of particle invariants. It has been
concluded that, thanks to the inclusion of the Carter constant, these
equilibrium solutions exhibit a non-isotropic character, which in turn
implies corresponding non-isotropic continuum fluid fields. To better
elucidate the issue, an asymptotic treatment of physical interest for the
kinetic equilibrium has been discussed, which has been referred to in the
paper as weak-anisotropy regime. This has allowed us to single out the
peculiar role of the Carter constant in providing the characteristic
non-isotropic features of the equilibrium solution in terms of a deviation
from an isotropic relativistic Maxwellian function.

Finally, the qualitative properties of the equilibrium stress-energy tensor
generated by these kinetic solutions have been discussed, proving that they
are generally of non-isotropic type. In particular, in the weak-anisotropy
regime, the analytical estimate has revealed the existence of a non-trivial
functional form for\ the stress-energy tensor in comparison with its
familiar representation holding for Maxwellian solutions. In fact, the
resulting tensor has been shown to be both non-isotropic as well as
generally non-diagonal when represented in terms of Boyer-Lindquist
coordinates. It is concluded that such features are peculiar of the
phase-space functional dependences carried by the Carter constant.

The kinetic theory developed here provides the appropriate theoretical
framework for the statistical description of both dynamical and
thermodynamical properties of collisionless neutral matter subject to
non-homogeneous gravitational fields in relativistic astrophysical regimes.
A relevant example of this type is represented by collisionless dark matter
halos which might arise in the surrounding of stellar-mass or
galactic-center black holes.

\bigskip

\textbf{Acknowledgments - }Work developed within the research projects of
the Czech Science Foundation GA\v{C}R grant No. 14-07753P (C.C.)\ and Albert
Einstein Center for Gravitation and Astrophysics, Czech Science Foundation
No. 14-37086G (Z.S.).

\end{document}